\documentstyle[a4,amssymb,psfig,12pt]{article}
\def\Frac#1#2{\frac{\displaystyle{#1}}{\displaystyle{#2}}}

\begin{document}

\rightline{MPI-PhT/96-93}

\begin{center}
{\Large {\bf Neutrino Oscillation and Magnetic Moment from $\nu -
e^{-}$ Elastic Scattering.}}
\end{center}

\vspace{0.5cm}

\begin{center}
{\Large {  J. Segura }}
\end{center}

\vspace{0.3cm}

\begin{center}
{\small {\it
Max-Planck-Institut f\"ur Physik. Werner-Heisenberg Institut. Munich.}}\\
{\small and}\\
 { \small  {\it Departamento de Ingenier\'{\i}a de Sistemas y
  Comunicaciones. Escuela
Polit\'ecnica Superior. Universidad de Alicante. E-03080 Alicante. Spain}}
\footnote{Postal Address. e-mail: segura@evalvx.ific.uv.es, jsegura@disc.ua.es}

\end{center}
\vspace{0.7cm}

\begin{abstract}

We discuss how the measurement of the $\bar{\nu}_{e}-e^{-}$ elastic
cross section at reactor energies can be used to extract new
information on the neutrino oscillation parameters. We also
consider the magnetic moment contribution and show how both effects
tend to cancel each other 
when the total cross section is measured; to achieve the
 separation of each of the effects, experiments capable of measuring angular
 and energy distributions with respect to the outgoing electron become
necessary. We study how their different energy and angle dependence
 enables such a separation; then, the sensitivity to
 magnetic moments, masses and mixings is discussed.
We also show how these experiments can be sensitive to the magnetic
moment of $\tau$ neutrinos via $\bar{\nu}_{e}\leftrightarrow \bar{\nu}_{\mu}$
oscillation.

\end{abstract}

Keywords: Neutrino-electron scattering, Neutrino Oscillations,
Neutrino magnetic moment, Reactor neutrinos.

PACs numbers: 13.10.+q;13.15.+g;14.60.Pq;13.40.Em. 

\newpage

\section{Introduction.}

Neutrino-electron elastic scattering is the most elementary 
purely weak leptonic scattering process; it involves very
well known particles ($e^{-}$) and neutrinos, whose
properties are the subject of a continuous theoretical and
experimental study\cite{Win}. Thus, this process appears as a crucial
probe for the study of neutrino properties.

Nevertheless, the smallness of the cross section seems a serious drawback
to obtain precision measurements from this process, specially when
low energy neutrinos from reactors are considered; from accelerators,
where higher statistics have been achieved, there exist measurements
of the weak couplings reaching a $5\%$ precision in the determination
of $sin^{2}\theta_{W}$\cite{Vil}. 
In reactors experiments, due to the lower statistics (less events are
 detected),
this achievement has been far from reach until now.

 Neutrino-electron elastic scattering experiments  have proven to
  be useful to establish
bounds on neutrino electromagnetic properties \cite{Rusos,Re,Yo0}, and, 
in particular,
on neutrino magnetic moments;
since the relative contribution of the magnetic moment interaction increases
with decreasing values of transferred momentum $Q^{2}$, electrons are the
the best target for such a purpose compared to any other, more massive,
 charged particle.  
The measurement of the total cross section in $\bar{\nu}_{e}-e^{-}$ 
elastic scattering at reactor experiments
give the most restrictive bounds on the magnetic moment of neutrinos ($\mu_{\nu}$)
since low energy neutrinos are available and low recoil electrons
can be detected ($\mu_{\bar{\nu}_{e}}<2.4\, 10^{-10}\mu_{B}$ 
 from $\bar{\nu}_{e} e^{-}\rightarrow \bar{\nu}_{e} e^{-}$\cite{Rusos}).
  New reactor experiments
   will be able to improve  such bound; for instance, the MUNU experiment
 will be
sensitive to magnetic moments as low as $2-3 \times 10^{-11}\mu_{B}$.
 In fact, the statistics
will be high enough and the control of the background good enough to
determine $sin^{2}\theta_{W}$ at the level of a $5\%$\cite{MUNU}. Apart from the
higher statistics, the most outstanding feature of the MUNU experiment
is the capability of measuring at some extent both the energy and recoil
angle of the electron thus being sensitive to the incoming neutrino energy.
Then, this new generation of detectors (MUNU, HELLAZ \cite{HEL}) will be able
 to measure at some extent distributions with respect to the energy $T$ and
  angle $\theta$ of recoil electrons.

Such experimental improvement allows the study of other properties of
neutrinos, apart from neutrino magnetic moment. As it was shown
in ref. \cite{Yo} the fact that the experiment is able to measure both the angle
and energy of the recoil electron, together with the cancellation of
the weak cross section $d\sigma^{\bar{\nu}_{e}}/dT$ ($T$ is the kinetic energy
of the recoil electron) for a neutrino energy
$E_{\nu}=m_{e}/4sin^{2}\theta_{W}$ and forward electrons gives rise to an
appearance-like experiment, where the study of neutrino oscillations is
available by measuring events around the {\it dynamical zero} \cite{Yo2}.
This novel kind of oscillation experiment was shown to be potentially 
sensitive
to values of masses and mixings not excluded by experiment yet.

In this work, we will estimate the sensitivity of present reactor 
experiments, taking MUNU as our reference, both for ú$\mu_{\nu}$ and
neutrino oscillations. We will show how a disappearance regime for the
oscillations is also available and that it is potentially sensitive 
to yet unbounded
values of the parameters; the oscillation contribution, however, tends
to cancel the $\mu_{\nu}$ contribution. Then, one should combine data
from both regimes to obtain independent bounds on each of the effects; 
angular (and energy) distribution measurements will be necessary since
one can not trust bounds on magnetic moment from total cross sections 
measurements at least one
disregards the possibility of neutrino oscillation. Besides, we will 
also show how, when $\bar{\nu}_{e}\leftrightarrow \bar{\nu}_{\tau}$  
oscillation is considered, the experiment is potentially sensitive to 
combinations of $\nu_{\tau}$ magnetic moment and mixings not excluded
by experiments so far.

\section{Appearance and disappearance regimes.}

The cross section  when  $\bar{\nu}_{e}\leftrightarrow \bar{\nu}_{X}$ 
($X=\mu,\tau$) oscillation  takes 
place within the distance $x$ from the reactor to the detector reads \cite{Yo}

\begin{equation}
\Frac{d\sigma}{dT}(x,E_{\nu},T)=\Frac{d\sigma^{\bar{\nu}_{e}}}
{dT}+{\displaystyle \sum_{i}}P_{\bar{\nu}_{e}\rightarrow\bar{\nu}_{i}}
(x)\left(\Frac{d\sigma^{\bar{\nu}_{\mu}}}
{dT}-\Frac{d\sigma^{\bar{\nu}_{e}}}
{dT}\right)
\end{equation}

\noindent
where we have used that $\sum_{i}P_{\bar{\nu}_{e}
\rightarrow\bar{\nu}_{i}}=1$ (we don't consider oscillation to sterile
 neutrinos) and $d\sigma_{\bar{\nu}_{\mu}}=
  d\sigma_{\bar{\nu}_{\tau}}$ in the Standard Model.
  
 To simplify the analysis, let us consider mixing of two generations
 ($\bar{\nu}_{e}\leftrightarrow \bar{\nu}_{\mu}$), then we should replace:

\begin{equation}
{\displaystyle \sum_{i}} P_{\bar{\nu}_{e}\rightarrow\bar{\nu}_{i}}
(x)\Rightarrow
P_{\bar{\nu}_{e}\rightarrow\bar{\nu}_{\mu}}
(x)=sin^{2}(2\phi)sin^{2}\left(\Frac{\Delta m^{2}x}{4E_{\nu}}\right)
\end{equation}

\noindent
where, as usual, $\Delta m^{2}$ is the difference of the squared masses
and $\phi$ the mixing angle.

Taking $E_{\nu}\simeq m_{e}/(4sin^{2}\theta_{W})$ and $\theta\simeq 0$
 ($\theta$ is
the electron scattering angle in the LAB frame) we would have
$d\sigma^{\bar{\nu}_{e}}<<d\sigma^{\bar{\nu}_{\mu}}$; then

\begin{equation}
\Frac{d\sigma}{dT}(x,E_{\nu},T)\simeq \Frac{d\sigma^{\bar{\nu}_{e}}}
{dT}+
 P_{\bar{\nu}_{e}\rightarrow\bar{\nu}_{\mu}}(x)
 \left(\Frac{d\sigma^{\bar{\nu}_{\mu}}}{dT}-
 \Frac { d\sigma^{\bar{\nu}_{e}} } {dT }  \right)>
 \Frac{d\sigma^{\bar{\nu}_{e}}}
{dT}
\end{equation}

\noindent
and we see how the kinematics have been chosen to perform an appearance-like
experiment in the sense that an excess of events would be detected
with respect to the standard model prediction. 

This is not the first time $\bar{\nu}_{e}-e^{-}$ has been considered in order
to study oscillations; nevertheless, ref. \cite{Yo} is
the first one to take advantage of the dynamical zero to perform an
appearance-like experiment. The previous analysis \cite{Mar}
 were based on the
measurement of the total cross section, in a given
 interval of recoil energies
$T$, integrating over the whole angular range  $\theta$ (not 
measured); considering this strategy at reactor energies,
 the $\bar{\nu}_{e}$
 contribution to (1) after
the integration over $T$ and all $\theta$ becomes around 2-3 times greater than
that from $\bar{\nu}_{\mu}$, due to the fact that for $\bar{\nu}_{e}$ we
have both charged and neutral currents and only neutral currents for
$\bar{\nu}_{\mu}$ (CC-NC interference is however negative, as shown
explicitly by the appearance of the {\it dynamical zero}). Then, far from
the dynamical zero where $d\sigma^{\bar{\nu}_{e}}>
d\sigma^{\bar{\nu}_{\mu}}$ (which covers most of the phase space) 
eq. (1) leads to

\begin{equation}
\Frac{d\sigma}{dT}(x,E_{\nu},T)\lesssim (1- P_{\bar{\nu}_{e}\rightarrow\bar{\nu}_{\mu}}
(x))
\Frac{d\sigma^{\bar{\nu}_{e}}}
{dT}<\Frac{d\sigma^{\bar{\nu}_{e}}}
{dT}
\end{equation} 

\noindent
and hence  the effect of oscillations would be to reduce
the number of detected events when one integrates over $T$ and all $\theta$;
in this case the kinematics have been chosen to perform a
 disappearance-like experiment. 
 
 In this type of disappearance-like experiment, contrary to the appearance
 case, one can not force one of the two cross section (for $\bar{\nu}_{\mu}$
  or $\bar{\nu}_{e}$) to be much smaller than the other one; then the
  relative contribution of the oscillation term will be smaller. But,
  on the other hand, we have the advantage of higher statistics since
  integrating  for all $\theta$. We will make use of both appearance
  and disappearance-like regimes to extract information on the oscillation
  parameters.
  
  \section{Estimative bounds on $\mu_{\nu}$ and the oscillation parameters.}
  
  Let us consider the observable:
  
  \begin{equation}
  R(\theta)=\Frac{N_{o}(\theta)}{N_{W}^{e}(\theta)}
  \end{equation}
  
  \noindent
  where $N_{o}$ is the number of events for
  electron angles lower than $\theta$ that would be detected by the experiment
  considering that oscillations occur and $N_{W}^{e}$ is the corresponding
  standard model prediction; when $\theta$ is small and the $T$-window 
  is located around $T\simeq 2m_{e}/3$ then $R>1$ (appearance) 
  while $R<1$ when we integrate
  $\forall \theta$ (disappearance).
  
  We will now estimate the bounds one could extract by measuring $R$,
  considering for the moment that the statistical error alone accounts
  for the precision of the experiment, which measures $N_{o}\pm \sqrt{N_{o}}$
  events.Then, a would-be exclusion plot 
  ($1\sigma$) could be derived by setting
  
  \begin{equation}
  \left|R-1\right|\equiv
  \left|\Frac{{\displaystyle \int } P (d\sigma^{\bar{\nu}_{\mu}} -
  d\sigma^{\bar{\nu}_{e}})}{N_{W}^{e}}\right|< \Frac{\sqrt{N_{o}}}{N_{W}^{e}}
  \equiv  F
  \end{equation}
  
  \noindent
  where  F accounts for the precision
  in the measurement of $R$.

  To make an estimative comparison between both regimes, let us integrate 
  events in the window $0.1<T <2 \,MeV$ for two different selection of angles:

  \begin{description}
  
  \item{a)} $\theta<0.3 \,\,\,rad$ 
  \item{b)} $\forall \,\,\theta$
  
  \end{description}
  
  We will take as $N_{o}$ ($N_{W}^{e}$ ,$N_{W}^{\mu}$)
  the total number of events in one year using as
   normalization the number
  of events that are expected to be detected by the MUNU experiment in
  the window $0.5<T<2.0\, MeV$; we take $x=20\, m$ (similar to the actual
  reactor-detector distance in the MUNU experiment).

  Using eq. (7) we get the following estimative limits:

\begin{center}
\begin{tabular}{|c||c|c|c|c|}
\hline \hline
  & $\Delta{m^{2}} (eV/c)^{2}>$  
  & $\Delta m^{2} (eV/c)^{2}>$
  & $sin^{2}(2\phi)<$
  & Statistical \\
  &  for $sin^{2}(2\phi)=1$
  &  for $sin^{2}(2\phi)=0.35$
  &  for $\Delta m^{2}>>$
  & Error\\
\hline  
  a) & $1.3\, 10^{-2}$ & $2.5\, 10^{-2}$ & $0.21$ & $9.0\%$\\
\hline 
  b) & $1.0\, 10^{-2}$ & $1.7\, 10^{-2}$ & $0.04$ & $1.3\%$\\
\hline
\end{tabular}
\end{center}

\noindent
 to be compared with those given by disappearance reactor experiments
 \cite{React}. Note
  how the disappearance regime is more sensitive to low
  values of $sin^{2}(2\phi)$ ($\left <P\right>\simeq \frac{1}{2} 
  sin^{2}(2\phi)$) than the appearance
  due to its higher statistics; this fact, contrary to the comparison
  between "standard" appearance and disappearance experiments, is
  related the fact that the statistics is higher in the
  disappearance regime for $\bar{\nu}_{e} e^{-}$ elastic scattering.
  On the other hand, for low $\Delta m^{2}$ we have
  
  \begin{equation}
  \left< P\right> \simeq k (\Delta m^{2})^{2}sin^{2}2\phi 
  \end{equation}
  
  \noindent
  with $k=\left<(x/4E_{\nu})^{2}\right>$  and one can check
  that higher values of $E_{\nu}$ enter in the integration
  b) compared to integration a); this explains why, being the bound    
  on $sin^{2}2\phi$ 
  for high $\Delta m^{2}$ greater
  in case a) the bounds for low $\Delta m^{2}$ are similar to those
  extracted from b). This particular discussion shows the general features
  of both regimes: similar sensitivity for low $\Delta m^{2}$ but better
  sensitivity of the disappearance regime to low $sin^{2}2\phi$
  (high $\Delta m^{2}$). 
 
 The limits extracted from the kinematical selection a) are
   worse than that for b); but, let us note that to extract all
     the potentiality of the {\it dynamical zero}, it will be necessary to
       consider narrower windows around \mbox{$T\simeq 2m_{e}/3$}
        (and $\theta=0$)
         where the cancellation in the standard cross section takes place. 
  
   In any case, this first estimation
 compels us to a closer look to the oscillation effect, since
 the estimative bounds presented are similar, for large mixing, to
 those obtained by C.C.-detection while we get lower bounds for
 the mixing at high $\Delta m^{2}$ for selection b). However,
  let us remind that, although the statistical error in case b) is quite small
   we should consider also the systematic error
  ($5\%$) coming mainly from the lack of precision in the knowledge of
  neutrino spectrum. Later, we will come back to this point. The sensitivity
  of the experiment MUNU, following \cite{MUNU}, is mainly dominated
  by this systematic uncertainty and changes very little as a function
  of the signal versus background ratio.

  Let us now estimate which would be the bound on $\mu_{\nu}$ that could be
  extracted from the measurement of events inside the region
  $0.1<T<2.0\, MeV$, $\forall \theta$. The signature of this contribution
  would be to increase the total number of expected events since now

  \begin{equation}
  \Frac{d\sigma}{dT}=\Frac{d\sigma_{e}^{W}}{dT}+\Frac{d\sigma_{e}^{M}}{dT}>
  \Frac{d\sigma_{e}^{W}}{dT} \,\,\,\, \forall\,\,\theta\, , \, T
  \end{equation}

  Then, considering only the statistical error, if the experiment measures
  $N\pm \sqrt{N}$ events, being $N=N_{W}^{e}+N_{mag}$,  one can set
  a bound on $\mu_{\nu}$; then, we get $\mu_{\nu}< 1.3\, 10^{-11}
   \mu_{B}$ ($1\sigma$) with the
  integration selection b). If one now integrates
  only for $\theta<0.3 \,$rad (a))one will have worse statistics ($9\%$
  compared to $1.3\%$) but,
  since the weak cross section becomes smaller, there is a better ratio
  signal  versus weak interaction an one can establish $\mu_{\nu}< 3.1\,
   10^{-11}
   \mu_{B}$. Then, at first sight, one could conclude that
   more stringent bounds, both for the oscillation parameters and for 
   $\mu_{\nu}$ are extracted when the total number of events for all angles
   are measured.
 
  However, comparing eqs. (4) and (8), we see that oscillation effects
  enter with the opposite sign when measuring events $\forall \,\theta$
  so that both effects could even cancel each other. Then, although
  the better statistics are achieved integrating over the whole range
  of energies and angles caution must be taken since the bounds,
  both for oscillation or neutrino magnetic moment, depend on the assumption
  of the absence of the other effect. On the other hand, if we now compare
  eqs. (3) and (8), we see how both effects tend to increase the total
  number of events. In fact, on the {\it dynamical zero} ANY extra
  contribution will be additive, if it appears, since the standard weak
  cross section is zero; then one could set bounds over one of
  such non-standard effects neglecting the rest. In particular,
   for $0.1<T<0.5 \,MeV$
  and $\theta <0.3\, rad$, under the same assumptions as before we get

\begin{center}
\begin{tabular}{||c||c|c|c|c|}
\hline
\hline 
  &  $\Delta{m^{2}} (eV/c)^{2}\geq$  
  & $\Delta m^{2} (eV/c)^{2}\geq$
  & $sin^{2}(2\phi)\leq$
  & Statistical\\
  &  for $sin^{2}(2\phi)=1$
  &  for $sin^{2}(2\phi)=0.35$
  &  for $\Delta m^{2}>>$
  &  Error\\
\hline  
  a) & $8.1\, 10^{-3}$ &  $1.5\, 10^{-2}$ & $0.35$ & $60\%$\\
\hline
  b) & $9.3\, 10^{-3}$   & $1.6\, 10^{-2}$   & 0.05  & $1.7\%$\\
 \hline 
\end{tabular}
\end{center}

\noindent
 while, in the same region, we get the bound $\mu_{\nu}
 < 2.8\, 10^{-11}\mu_{B}$ for a) and
  $\mu_{\nu}
   < 1.2\, 10^{-11}\mu_{B}$ for b).  With this selection in the
 $T$-window we obtain, with a much higher statistical error, bounds on 
 oscillations parameters
 somewhat better than those from disappearance-like for high
 of $sin^{2}2\phi$  and similar bounds for $\mu_{\nu}$ 
  but worse bounds for $sin^{2}2\phi$ at
 high $\Delta m^{2}$ . In any case,
 it seems that both appearance and disappearance regimes can potentially
 give relevant information on the oscillation effect and that similar
 bounds on $\mu_{\nu}$ can be given with both kinds of integration region;
 unfortunately, as commented before, the disappearance region shows
 a clear disadvantage: $\mu_{\nu}$ and oscillation effects tend to cancel;
 this fact is illustrated in figure 1 where the $\sigma$-deviation of the
 observable R, both for oscillation and magnetic moment, with respect
 to $1$ is plotted as a function of the angle of integration; in this figure
 the parameters are fixed to one of the bounds extracted from the integration
 $\forall \theta$ (b)).

 Then, as a summary of our previous estimations, and to fix ideas, let us
 underline some facts:
 
 \begin{description}

 \item[1.-] Both regimes are sensitive to $\bar{\nu}_{e}\leftrightarrow
 \bar{\nu}_{x}$ where $\bar{\nu}_{x}$ stands for any non-sterile neutrino;
  there is no production threshold for the detection.
 
 \item[2.-] Similar bounds for large $sin^{2}2\phi$ are extracted from
 appearance and disappearance regimes, but disappearance gives better 
 bounds for small mixing (neglecting $\mu_{\nu}$ interaction). The 
 statistical error is much larger in the appearance regime.

 \item[2.-] Disregarding oscillations, $\mu_{\nu}$ is better measured
 integrating $\forall \theta$.
 
 \item[3.-] Disappearance-like contributions could mask $\mu_{\nu}$
  interaction when an integration $\forall \theta$ is performed.
  
  \item[4.-] Any non-standard effect would increase the number of detected
  events around the dynamical zero ($T\simeq 2m_{e}/3$ and $\theta =0$).
 
 \end{description}

  \section{Disentangling the magnetic moment effect from oscillation.}

     Only if an experiment is able to measure both the recoil angle
     of the electron $\theta$ and its energy $T$ one can safely separate
     both effects, since, as explained, though they tend to cancel each
     other when total cross sections are measured, the angular (and energy)
    distributions for oscillation and $\mu_{\nu}$-interaction are different
 (the $\mu_{\nu}$ term is always additive and oscillations add near the
 {\it dynamical zero} but they subtract far away from it); furthermore,
  one can conclude from the previous
 estimations that both regions (far away and close to the {\it dynamical zero})
 are sensitive to oscillations and magnetic moment. The one could fit
 the shape of the distributions to separate both effects; a deviation
 from the expected $\theta$ dependence would be mainly a signal of
 oscillation while $\mu_{\nu}$ would be observed as an overall excess of
  events. One can also construct convenient
  observables to achieve such a separation; for instance, and as an
  illustration, let us consider
   the  quantity:

 \begin{equation}
 {\cal{O}}_{o}=\Frac{N_{o}(\theta <\theta_{0})}
 {N_{o}(\theta >\theta_{0})}\left/\Frac{N_{W}(\theta <\theta_{0})}
 {N_{W}(\theta >\theta_{0})}\right.
 \end{equation}

 \noindent 
 where, given a window in $T$, $N(\theta <\theta_{0})$ is the number
 of observed events $N_{o}$ and of expected events from the S.M. predictions
 ($N_{W}$) that would be detected for angles lower than $\theta_{0}$;
 $N(\theta >\theta_{0})$ is the corresponding number of events for
 angles higher than $\theta_{0}$.
 
 Considering an observable such like ${\cal{O}}_{o}$ will have several
 benefits: first, we can take advantage of both the apparition
 ($N(\theta<\theta_{0})$) and disappearance regimes
  \mbox{($N(\theta>\theta_{0})$)} for the oscillation contribution;
 second, since we are integrating in the same window of $T$ both
 for $\theta <\theta_{0}$ and  $\theta >\theta_{0}$ similar
 neutrino energies appear, thus canceling neutrino
 spectrum uncertainties partially and total flux uncertainty completely; 
 and third, due to their different angular dependence
 the observable will be able to separate oscillation contributions
 from magnetic moment contributions.

  With all this arguments in mind, let us check what is the $\sigma$
  deviation of ${\cal{O}}$ from $1$ (value without non-standard effects) for
  different values of $\theta_{0}$ and different selections in the
  window in $T$; since we expect some dependence $ {\cal{O}}(\theta_{0})$
  we will also estimate the error in the measurement of
  $\theta$, taking ${\cal{E}}(\theta)=0.05\, rad$ \cite{MUNU};
  we will consider that the precision is given mainly by the
  statistical error and by the uncertainty in the determination of $\theta$
  and thus in the actual value of the observable
   ${\cal{O}}_{o}(\theta_{0})$ (systematic error);
  we sum in quadrature both uncertainties. 
  Fig. (2) shows the sigma deviations of ${\cal{O}}_{osc}$ for different
  selection of $T$-windows and oscillation parameters. In the same figure,
  we also plot the $\sigma$-deviation of the observable
  for electromagnetic interaction taking $\mu_{\nu}=2.3\, 10^{-11}\mu_{B}$

  From fig. (2) we observe that the region of low $\theta \sim 0.3\, rad$
  is mainly sensitive to oscillation; in this region the observable ${\cal{O}}$
   is not
  much sensitive to $\mu_{\nu}$; also, let us notice that similar bounds
  for high $sin^{2}2\phi$ to those from the measurement of $R$ can be
  extracted (both appearance and disappearance were equally sensitive to
  this values) while the bounds on $sin^{2}2\phi$ at high 
  $\Delta m^{2}$ are between those extracted from 
  the appearance and disappearance regimes, improving the first ones; 
  for instance, in the window $0.1<T<0.5 \, MeV$ and for $sin^{2}(2\phi)=1$
  one gets $\Delta m^{2}< 10^{-2} eV^{2}$ at $\theta_{0}=0.3\, rad$
  with a $40\%$ error in the measurement of ${\cal{O}}$ (curve 1, dashed)
   while for the window $0.1<T<2.0 \,
  MeV$  and large $\Delta m^{2}$ one can set
  $sin^{2}(2\phi) <0.15$ with a $10 \%$ error (curve 2, solid). 
 
   On the other hand, for large
   $\theta\sim 1 rad$ we have a better sensitivity to $\mu_{\nu}$, which
   also lies between the bounds extracted from 
    the selections a) and b) for the angular region ($\mu_{\nu}<
    2.3\, 10^{-10}\mu_{B}$ for $0.1<T<2.0\, MeV$ with a $\,6\%$
    error. Then,
   we see that there are two regions and each of them is
    mainly sensitive to one of
   the effects considered.  Besides, for small angles, the mag. moment
    effect is of the same
   sign as the oscillation one so that there
    is no risk of cancellation between them. Notice also the dependence
    of the effects on the selection of $T-window$; mag. moment. contribution
    and oscillation for large $\Delta m^{2}$ are better measured when
    the $T$-window is large (since in this case the most important
     thing is to get
    as high statistics as possible, as it was shown previously) while
     for large $sin^{2}2\phi$ and small $\Delta m^{2}$ it is more
     convenient to chose a narrower window around the dynamical zero
     (in our case $0.1<T<0.5 MeV$). Then, this is a tunable
     experiment which can prospect different regions of the parameter
     space by choosing different values of the angle and recoil energy.
 
 \section{$\bar{\nu}_{\tau}$ electromagnetic properties from
  $\bar{\nu}_{e}-e^{-}$ elastic scattering.}
 
 Finally, it is interesting to note that, in case oscillations
 take place, any extra interaction of the new flavors 
 originated from oscillations, could affect the value of the
 elastic cross section; consider, for instance, that $\bar{\nu}_{e}$
 oscillate to $\bar{\nu}_{\tau}$ and that the tau neutrino has a large magnetic
 moment  (the best lab. bound is $4\, 10^{-6}\mu_{B}$
 \cite{mutau}). To simplify the analysis, let us consider
 that the magnetic moment cross section for the tau neutrino
 is much larger than the weak cross sections (which is granted for
 $\mu_{\nu}>10^{-9}\mu_{B}$) and that $\tau$-neutrino mass effects are
  negligible in the cross section; then, we would expect an excess to respect to the
 standard model prediction and one can write
 
 \begin{equation}
 \Frac{d\sigma}{dT}\simeq \Frac{d\sigma^{\bar{\nu}_{e}}}{dT}+
 \mu^{2}P(x)\Frac{\pi \alpha^{2}}{m_{e}^{2}}\Frac{
 1-T/E_{\nu}}{T}\,\,\,;\,\, \mu=\mu_{\bar{\nu}_{\tau}}/\mu_{B}
 \end{equation}
 
 \noindent
 where the terms $P(x)d\sigma^{\bar{\nu}_{e}}$ and $P(x)d\sigma^
 {\bar{\nu}_{\tau}}$ have been neglected. Then one gets the following
 $1\sigma$ bounds from the observable ${\cal{O}}$:
 
 \begin{description}
 
 \item[--]  $\chi^{2} sin^{2}(2\phi) < 0.1$ at large $\Delta m^{2}$ 
 for $0.1<T<2.0\, MeV$, $\theta\sim
 1\,rad$; total error $\sim 5\%$.
 
 \item[--] $\chi^{2} sin^{2}(2\phi) (\Delta m^{2})^{2} < 2\,10^{-5} (eV^4) $
  at large
 $sin^{2}(2\phi)$ for $0.1<T<0.5\, MeV$, $\theta\sim 0.4\,rad$; total
 error $\sim 20\%$. 
 
 \end{description}
 
 \noindent 
 where $\chi=\mu_{\bar{\nu}_{\tau}}/10^{-10}\mu_{B}$.
 
  This kind of bounds,
 although relating unknown parameters, would explore combinations of values
 which are not excluded yet; or, in other words, the elastic
  neutrino-electron cross section is sensitive to non standard neutrino physics 
 in still admissible scenarios.

   \section{Conclusions.}
 
    We should stress again that the measurement
    of the total cross section in present reactor experiments 
    is potentially sensitive both to values of $\mu_{\nu}$ and the oscillation
    parameters not excluded yet. By measuring the total cross section
     one can not set
    bounds on $\mu_{\nu}$ neglecting oscillation since both effects tend
    to cancel. Experiments able to measure energy and angle
    of recoil electrons are needed; only then one can safely separate
    both effects and set independent bounds on each of them. Furthermore,
    the elastic neutrino-electron cross section is sensitive to non-standard
    neutrino interactions induced by oscillation, as is the case of
    magnetic moment interaction of $\bar{\nu}_{\tau}$ ($\bar{\nu}_{\mu}$), for
     still
    available values of the parameters. Therefore, this process promises
    a better understanding of neutrino dynamics.

\vspace{0.5cm}

{\bf Acknowledgements.}

 The author wishes to acknowledge the support from {\it Fundaci\'on 
 Ram\'on Areces }
for his postdoctoral fellowship. I also wish to thank 
  G. Raffelt
for his hospitality during my stay at M.P.I. f\"ur Physik-Munich and
M.C. Gonz\'alez-Garc\'{\i}a, L.M. Garc\'{\i}a-Raffi and A. Gil for
useful discussions.

\newpage

{\bf \underline{Figure captions:}}

\vspace{0.5cm}

{\bf Figure 1:} $\sigma$-deviation of the observable $R(\theta)$ from 1
  for  $0.1<T<0.5 \, MeV$ (dashed) and $0.1<T<2.0$ (solid).
  $\Delta m^{2}=10^{-2} eV^{2}$ and $sin^{2}2\phi=1$ for curves {\bf 1}; 
 $\Delta m^{2}=1 eV^{2}$ and $sin^{2}2\phi=0.04$ for {\bf 2}; 
  $\mu_{\nu}=1.2\, 10^{-11}\mu_{B}$ for {\bf 3}.\\

{\bf  Figure 2:} Sigma-deviation 
of the observable ${\cal{O}}_{o}(\theta_{0})$ from $1$  
 for $0.1<T<0.5 \, MeV$ (dashed) and $0.1<T<2.0 \,MeV$ (solid).
   $\Delta m^{2}= 10^{-2} eV^{2}$ and $sin^{2}2\phi=1$ for curves {\bf 1};
    $\Delta m^{2}=1 eV^{2}$ and $sin^{2}2\phi=0.15$ for {\bf 2};
      $\mu_{\nu}=2.3\, 10^{-11}\mu_{B}$ for {\bf 3}.

\newpage

\vskip 0.2cm
\centerline{\protect\hbox{\psfig{file=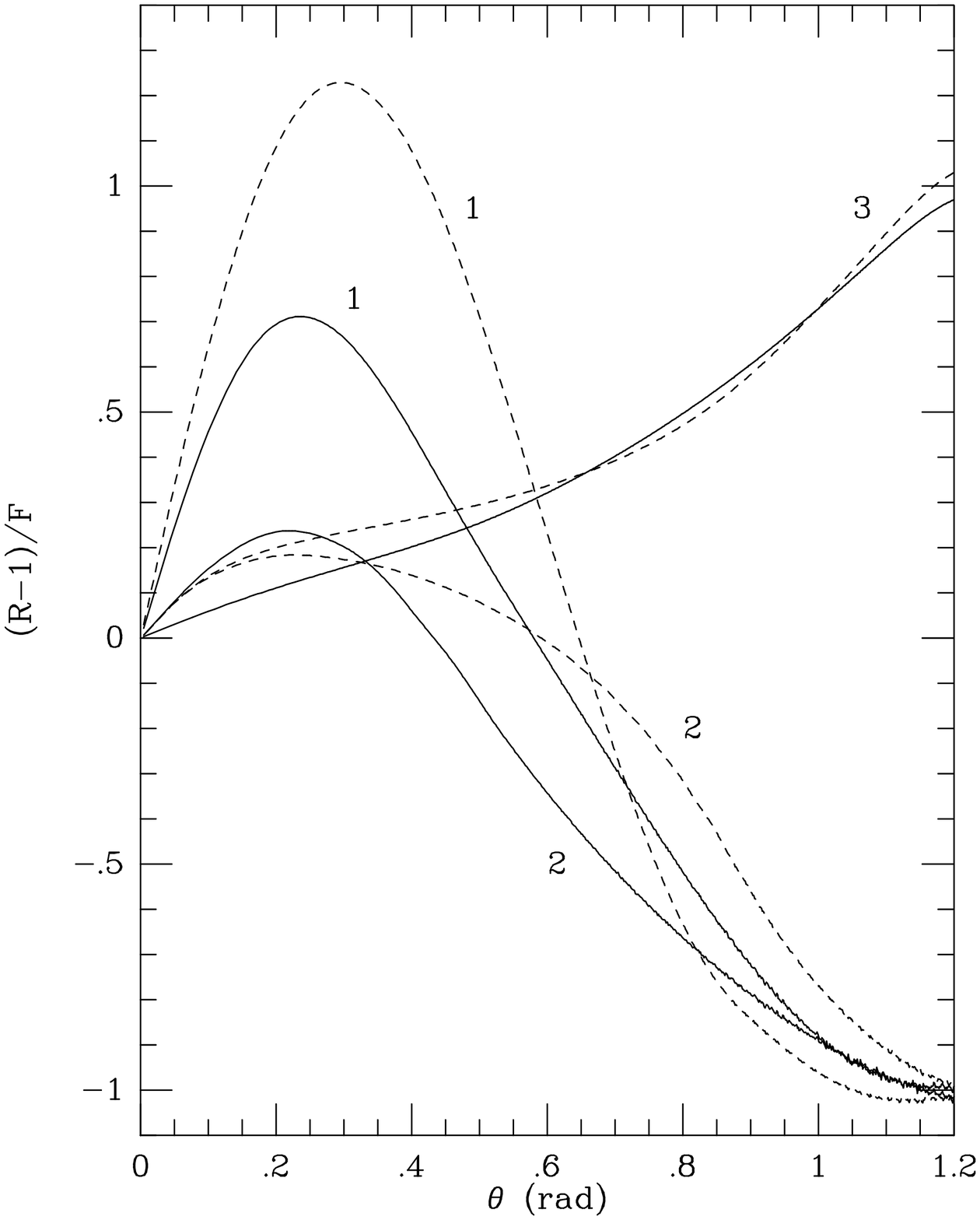,width=17cm}}}
\vskip 0.2cm
\flushright
{\bf Figure 1}

\newpage

\vskip 0.2cm
\centerline{\protect\hbox{\psfig{file=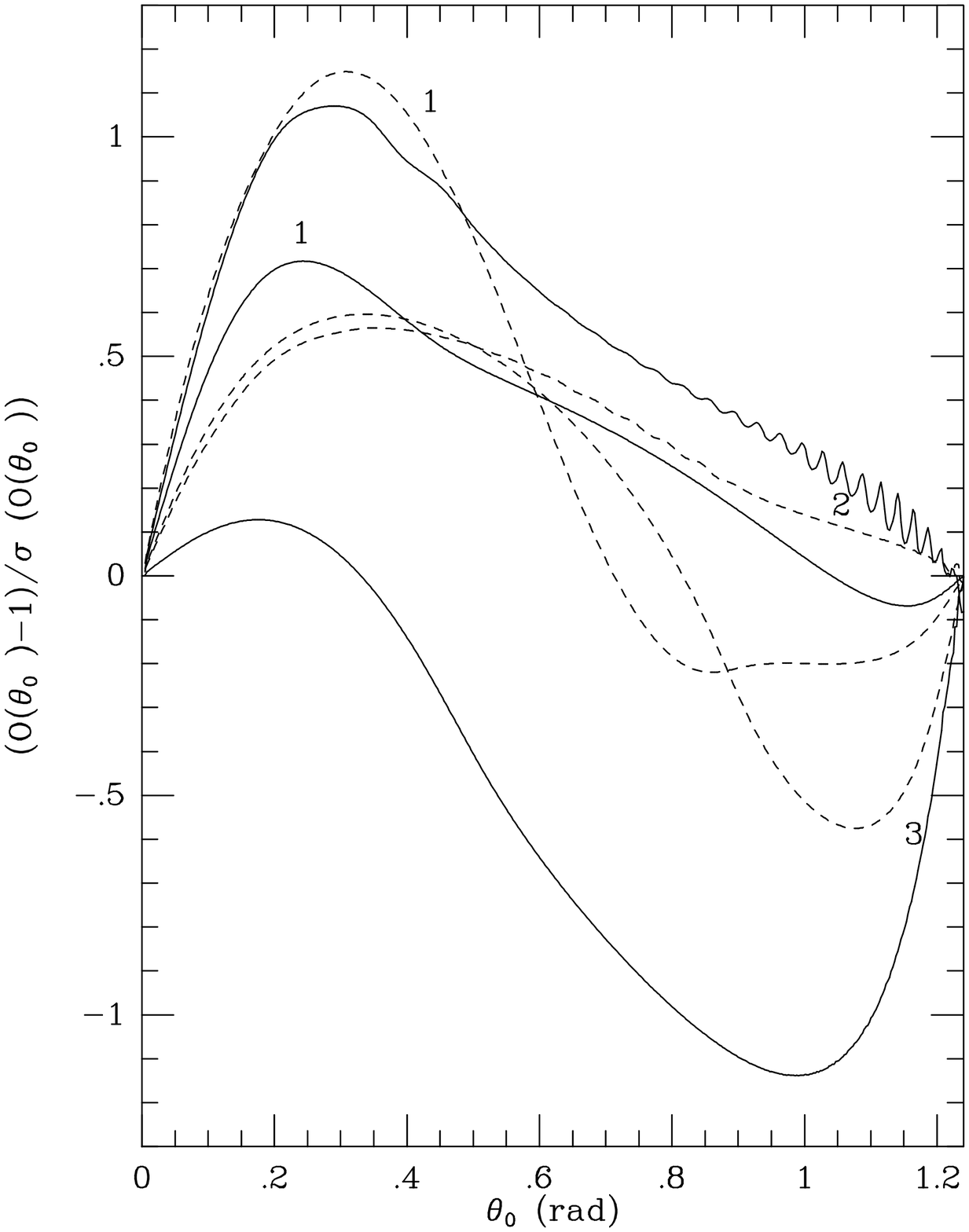,width=16.5cm}}}
\vskip 0.2cm
\flushright
{\bf Figure 2}


\vspace{0.2cm}


\begin{thebibliography}{99}

\bibitem{Win} K. Winter, Lepton-Photon Symp.(1995)569.\\
              G. G. Raffelt, Blois 1992, Proceedings, Particle Astrophysics
               99-110. 

\bibitem{Vil} P. Vilain et al., Phys. Lett B335(1994)246.

\bibitem{Rusos} G.S. Vidyakin et al., JETP Lett. 49(1989)740; JETP Lett. 
                55 (1992) 206.

\bibitem{Re} D.A. Krakauer et al., Phys. Lett. B252(1990) 177;\\
             P. Vilain et al., Phys. Lett. B345(1995)115.

  
\bibitem{Yo0} J. Bernab\'eu, S.M. Bilenky, F.J. Botella y J. Segura,
             Nucl. Phys. B246(1994)434.


\bibitem{MUNU}  C. Broggini et al., LNGS-92/47 (MUNU proposal).

\bibitem{HEL} F. Arzarello et al., Preprint CERN-LAA-94-19. 

\bibitem{Yo}  J. Segura, J. Bernab\'eu, F.J. Botella y J.A. Pe\~{n}arrocha,
  Phys. Lett. B335(1994)93.
  
\bibitem{Yo2}  J. Segura, J. Bernab\'eu, F.J. Botella y J. Pe\~{n}arrocha,
    Phys. Rev. D49(1994)1633.
    
\bibitem{Mar} W.J. Marciano, Phys. Rev. D36(1987)2859;\\
              S.P. Rosen, B. Kayser, Phys. Rev. D23(1981)669;\\
              L.A. Ahrens et al., Phys. Rev. D31(1985)2732.\\
              B. Halls, B.H.J. McKellar, Phys. Rev. D24(1981) 1785.

\bibitem{React} B. Achkar et al., Nucl. Phys. B434(1995)503;\\
                G. Zacek et al., Phys. Rev. D34(1986)2621;\\
                G.S. Vidyakin et al., JETP Lett. 59(1984)364.

\bibitem{mutau} H. Grotch, R. Robinet, Z. Phys. C39(1988) 563;\\
                T.M. Gould, I.Z. Rothstein, Phys. Lett. B333(1994)545.

\end{thebibliography}
\end{document}